\documentclass[12pt,preprint]{article}
\usepackage{scicite}
\usepackage{times}
\usepackage{graphicx,morefloats}
\usepackage{subfigure} 
\usepackage{color,soul}
\usepackage{caption}
\usepackage{bm}
\usepackage[]{multibib}
\newcites{SM}{References Supplementary Materials}

\newenvironment{sciabstract}{%
\begin{quote} \bf Complex and correlated quantum systems with promise for new functionality often involve entwined electronic degrees of freedom.
In such materials, highly unusual properties emerge
and could be the result of electron localization.
Here, a cubic heavy fermion metal governed by spins and orbitals is chosen as a model system for this physics. Its properties are found to originate from surprisingly simple low-energy behavior, with two distinct localization transitions driven by a single degree of freedom at a time. This result is unexpected, but we are able to understand it by advancing the notion of sequential destruction of an SU(4) spin-orbital-coupled Kondo entanglement. Our results implicate electron localization as a unified framework for strongly correlated materials and suggest ways to exploit multiple degrees of freedom for quantum engineering.
\end{quote}}

\title{Sequential localization of a complex electron fluid} 

\author{
\large{V.\ Martelli$^{1,+,\oplus}$, A.\ Cai$^{2,\oplus}$, E.\ M.\ Nica$^{2,\times}$, M.\ Taupin$^1$, A.\ Prokofiev$^1$,}\\
\large{C.-C.~Liu$^2$, H.-H.\ Lai$^2$, R.\
Yu$^{2,3}$, K. Ingersent$^4$, R.~K\"uchler$^5$, A.\ M.\ Strydom$^6$,}\\
\large{D.\
Geiger$^1$, J.\ Haenel$^1$, J.~Larrea$^{1,+}$, Q.\ Si$^{2,\ast}$, and
S.~Paschen$^{1,2,\ast}$}\\
\\[-0.5cm]
\normalsize{$^1$Institute of Solid State Physics, Vienna University of Technology, 1040
Vienna, Austria}\\
\normalsize{$^2$Department of Physics and Astronomy, Rice Center for Quantum Materials,}\\
\normalsize{Rice University, Houston, TX 77005, USA}\\
\normalsize{$^3$Department of Physics, Renmin University of China, Beijing 100872, China}\\
\normalsize{$^4$Department of Physics, University of Florida, Gainesville, Florida 32611-8440, USA}\\ 
\normalsize{$^5$Max Planck Institute for Chemical Physics of Solids, N\"othnitzer Str.\ 40,}\\
\normalsize{01187 Dresden, Germany}\\
\normalsize{$^6$Highly Correlated Matter Research Group, Physics Department,}\\ \normalsize{University of Johannesburg, Auckland Park 2006, South
Africa}\\[-0.3cm]
}

\date{}

\begin{document}
\baselineskip24pt

\maketitle 

\begin{sciabstract}
\end{sciabstract}

\noindent{\bf Significance statement:} Many of the most fascinating and actively
investigated materials classes host strongly correlated electrons. Their
understanding is challenging because the strong correlations cause entwining of
multiple degrees of freedom of an electron, such as spin, orbital, and charge.
This complexity is ubiquitous and underlies many of the rich properties. The
question then is whether there are universal organizing principles that provide
simplicity to the description. Here, by studying a prototype material with
entwined spin and orbital degrees of freedom and a theoretical model pertinent
to it, we have demonstrated correlation-driven electron
localization-delocalization as such a principle. It happens sequentially,
involving a single quantum number at a time, thus deciphering the roles of the
individual degrees of freedom.

\vspace{1cm}


\noindent $^{\ast}$Corresponding author. qmsi@rice.edu, paschen@ifp.tuwien.ac.at

\noindent $^{\oplus}$V.M. and A.C. contributed equally to this work.

\noindent \hspace{0.2cm} Present addresses: $^+$University of S$\mathrm{\tilde{a}}$o Paulo, S$\mathrm{\tilde{a}}$o Paulo, Brazil;

\hspace{2.9cm}$^{\times}$Department of Physics and Astronomy and Quantum Matter Institute,

\hspace{3.1cm}
University of British Columbia, Vancouver, B.C., V6T 1Z1, Canada.

\noindent{INTRODUCTION}\\
Strongly correlated electron systems represent a vibrant frontier in modern
condensed matter physics. They often contain multiple degrees of freedom, which
may be harnessed for future applications in electronic devices. One famous
example is the manganites, in which both spin and orbital degrees of freedom
play an important role \cite{Tok00.1}. Others are the iron-based superconductors
\cite{Si16.1} and fullerides \cite{Tak09.1}. In the cuprates, charge order
emerges and interplays with the spin degrees of freedom to influence their
low-energy properties \cite{Bad16.1,Ram15.1}. Even in magic-angle graphene, the
physics likely depends on both the spin and valley degrees of freedom
\cite{Cao18.1}.  These systems display a rich variety of exotic properties at
low 
energies~\cite{Bad16.1,Ram15.1,Bal03.1,Par08.1,Sch00.1,Pas04.1,Sch16.1,Oik15.1,Cao18.1}.
Finding simplicity out of this complexity is a central goal of the field. An
emerging notion is that electron localization may be an organizing principle
that can accomplish this goal \cite{Si13.1}.\\

\noindent{RESULTS}\\
We have chosen heavy fermion materials as setting for our study because they can
be readily tuned to localization transitions and display sharp features thereof.
The $f$ electron's spin in a heavy fermion compound corresponds to a
well-defined local degree of freedom. At the same time, it is still sufficiently
coupled to the conduction electrons so that its behavior can be probed through
the latter. In the ground state, Kondo entanglement generally leads to the
formation of a many-body spin singlet between the local moment and conduction
electrons. Electronic localization of this electron fluid can then be realized
as function of a non-thermal control parameter
\cite{Sch00.1,Pas04.1,Par08.1,Fri10.2,Cus12.1,Luo14.1,Sch16.1,Wu16.2,Pro18.1x},
and has been  understood in terms of the destruction of Kondo entanglement
\cite{Si01.1,Col01.1,Sen04.1,Cai19.1x}. The accompanying strange-metal behavior,
as well as the onset of magnetic ordering of the liberated spins, and
unconventional superconductivity are prominent features
\cite{Sch00.1,Pas04.1,Par08.1,Fri10.2,Cus12.1,Luo14.1,Sch16.1,Wu16.2,Pro18.1x}
that make this transition both readily observable and broadly important.

\begin{figure}[t]
\centering
\subfigure[]{\includegraphics*[width=0.45\textwidth]{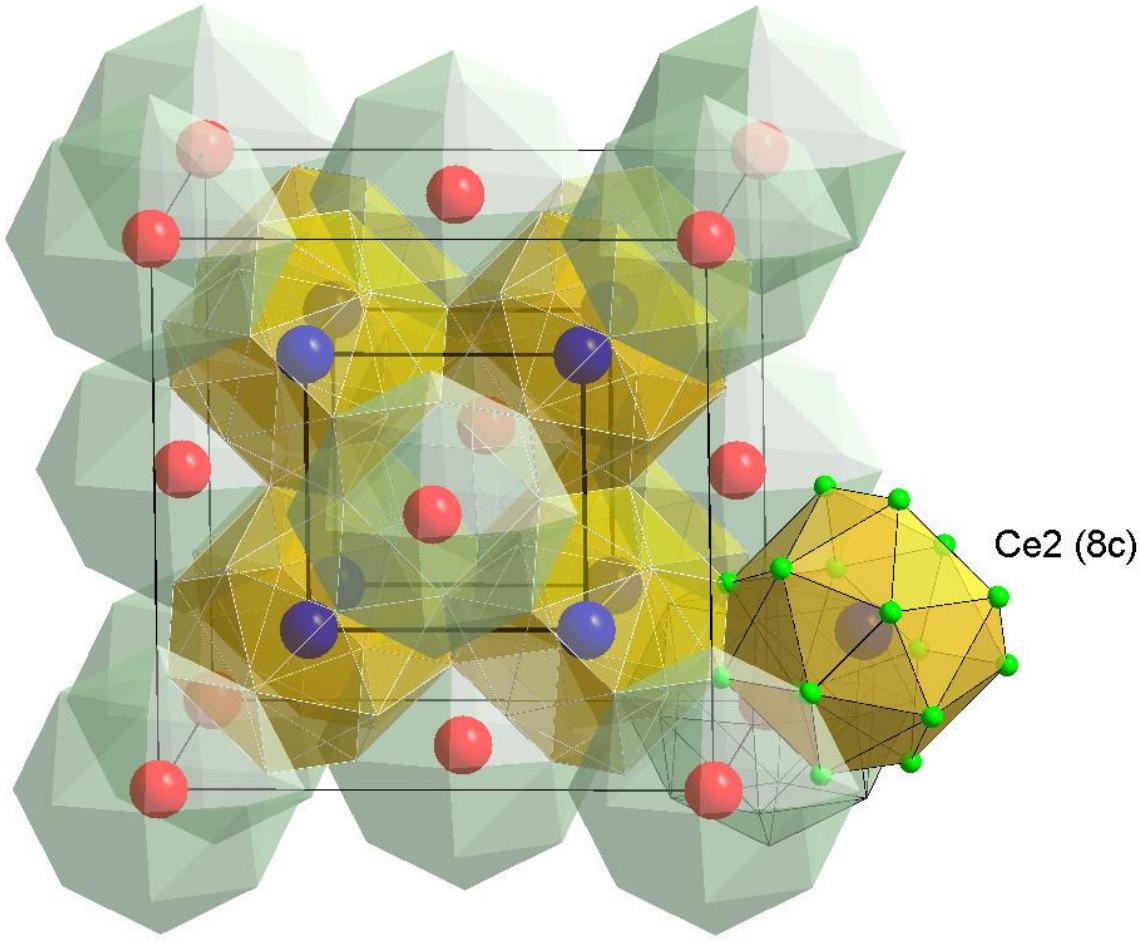}}\hspace{0.5cm}
\subfigure[]{\includegraphics*[width=0.45\textwidth]{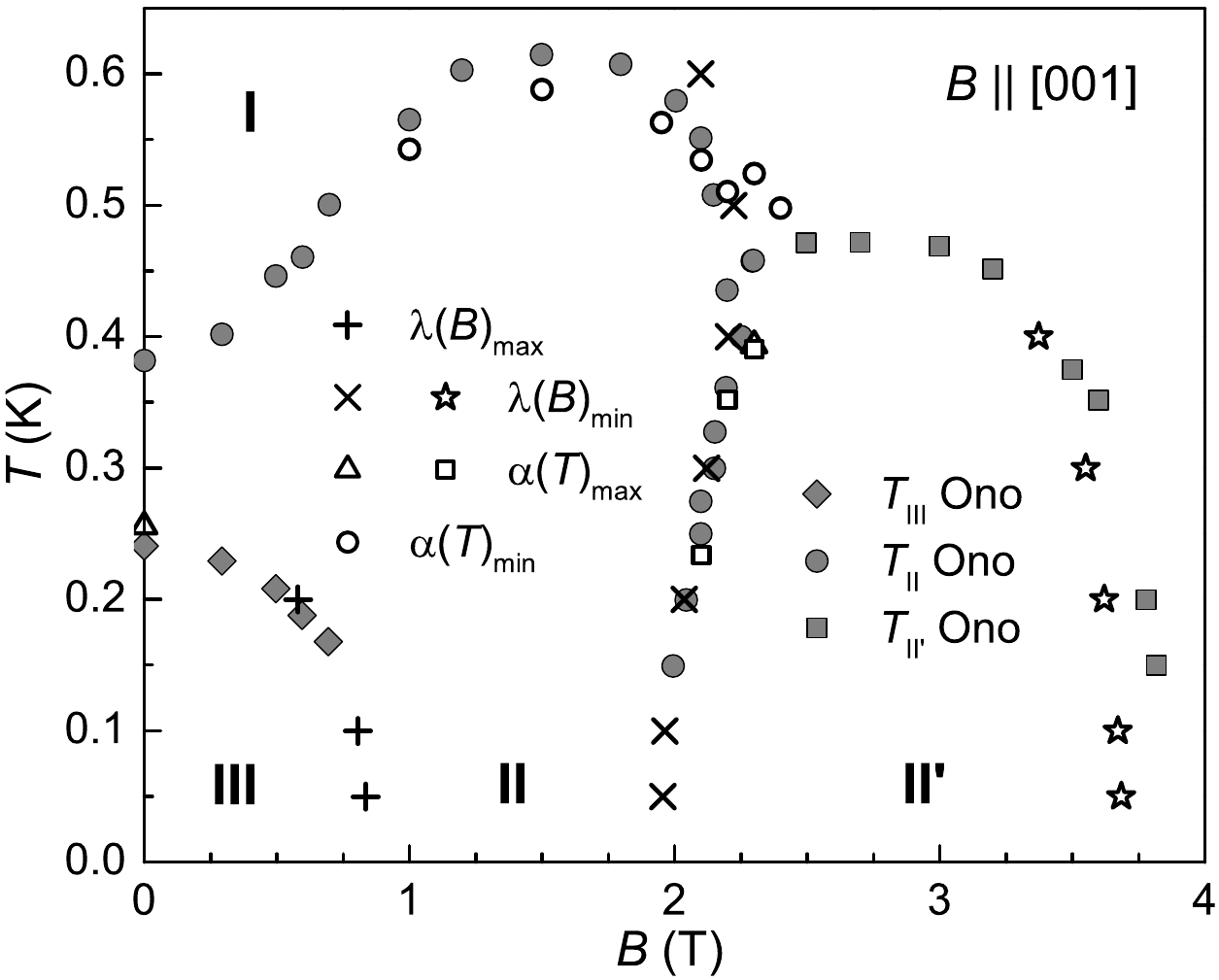}}
\caption{\label{fig1}~{\bf Crystal structure and ordered phases of the heavy
fermion compound Ce$_3$Pd$_{20}$Si$_6$.} (a) Cubic crystal structure of
space group $Fm\bar{3}m$ (Ref.\citenum{Gri94.1}) with the two Ce sites $4a$
(Ce1, red) and $8c$ (Ce2, blue), both with cubic point symmetry, forming a
face-centered cubic lattice of lattice parameter $a = 12.275$\,\AA\
(Ref.\citenum{Pro09.1}) and a simple cubic lattice of half the lattice
parameter, respectively. The polyhedra around Ce1 are made up of Si and Pd,
those around Ce2 of Pd only. This structure persists down to at least 40~mK, as
shown by high-resolution neutron diffraction measurements \cite{Dee10.1}. (b) Temperature-magnetic field phase diagram for fields $B=\mu_0 H$ applied
along $[0\,0\,1]$. The phase boundaries are determined from specific heat data
by Ono et al.\ \cite{Ono13.1} ($T_{\rm III}$\,Ono, $T_{\rm II}$\,Ono, and
$T_{\rm II'}$\,Ono refer to anomalies upon entering phase III, II, and II',
respectively), and our magnetostriction ($\lambda(B)_{\rm{max}}$ and
$\lambda(B)_{\rm{min}}$ mark the positions of the maxima and minima in
$\lambda(B)$, fig.\,S1B) and thermal expansion data ($\alpha(B)_{\rm{max}}$ and $\alpha(B)_{\rm{min}}$ mark the positions of the maxima and minima of $\alpha(T)$, fig.\,S1C). Phase I is paramagnetic, the order of phases II and III was identified as AFQ and AFM order of moments on the $8c$ cite, respectively; the nature of the order of phase II' remains to be identified \cite{Por16.1}. Neutron scattering has not detected any order associated with the $4a$ site \cite{Por16.1}. Phase III is isotropic with respect to the field direction, but phase II extends to fields above 10\,T for
fields along $[1\,1\,0]$ and $[1\,1\,1]$ (Ref.\citenum{Ono13.1}). Thus, it is
advantageous to study $B || [0\,0\,1]$, as done in the present work.
} 
\end{figure}

To explore the intricate interplay of multiple quantum numbers in this setting,
a local degree of freedom in addition to the electron's spin should come into
play. The simplest such case in heavy fermion systems may arise in cubic
Ce-based compounds. Due to strong intraatomic spin-orbit coupling, the spin and
orbital degrees of freedom of the Ce $4f^1$ electron are described in terms of
the total angular momentum $\bm{J}$, that encompasses both spins (dipoles) and
higher multipolar moments.  Ce- and Yb-based heavy fermion materials often have
crystalline symmetries lower than cubic.  In that case, the lowest crystal
electric field (CEF) level would be a Kramers doublet.  In the cubic case,
however, symmetry allows for CEF levels with higher degeneracy,  such as the
four-fold $\Gamma_8$ level, both in the case of the [Xe]$4f^1$ wavefunction  of
a Ce$^{+3}$ ion (for the total angular momentum $J=5/2$) or the [Xe]$4f^{13}$
wavefunction of a Yb$^{+3}$ ion  (for $J=7/2$). When this level is the lowest in
energy, we end up with one $f$-electron (or hole  in the Yb-based systems)
occupying a four-fold-degenerate local level,  which  can be  characterized by
spin {\em and} orbital quantum numbers \cite{Shi97.1}. This is indeed the case
in the intermetallic compound Ce$_3$Pd$_{20}$Si$_6$ (Fig.\,\ref{fig1}(a), see
also Section S4). At zero field, it is at first the quadrupolar moments that
order into an antiferroquadrupolar (AFQ) phase with ordering wave vector
$[1\,1\,1]$ at $T_{\rm{Q}} \sim 0.4$\,K; with further decreasing temperature,
the dipolar (magnetic) moments undergo antiferromagnetic (AFM) ordering, with
the ordering wave vector $[0\,0\,0.8]$ at $T_{\rm{N}} \sim 0.25$\,K, as shown by
recent neutron scattering experiments \cite{Por16.1}. Both orders are due to Ce
atoms on the crystallographic $8c$ site.

As typical for heavy fermion systems, the many-body ground state is readily
tunable by external parameters such as magnetic field. Previous work on
Ce$_3$Pd$_{20}$Si$_6$ polycrystals \cite{Cus12.1} indeed revealed the
suppression of $T_{\rm{N}}$ at a critical field $B_{\rm{N}}$. Quantum
criticality was revealed by electrical resistivity and specific heat
measurements; the temperature dependencies were found to be different from the
expectations \cite{Ste01.1} of the conventional theory based on order parameter
fluctuations. Measurements of magnetotransport revealed a jump of the Hall
coefficient and magetoresistance in the zero-temperature limit across
$B_{\rm{N}}$, which implicates a sudden reconstruction from large to small Fermi
surface with decreasing field, as expected for a localization transition
of Kondo destruction type \cite{Cus12.1}. When single crystals became available
(see also Section S1), the phase diagram was mapped out for different field
orientations \cite{Ono13.1}. The AFM transition is suppressed isotropically,
implying that the quantum critical behavior at $B_{\rm{N}}$ observed in
polycrystals captures the behavior of the single crystals. By contrast, the AFQ
transition is suppressed anisotropically \cite{Ono13.1,Por16.1}. The study of
the interplay between spin and orbital degrees of freedom thus requires
measurements on single crystals, which we carry out in the present work.

We chose to apply magnetic field along the crystallographic $[0\,0\,1]$
direction, which suppresses the AFQ phase at a relatively small field
$B_{\rm{Q}}$ (see Section S2). The temperature-magnetic field phase
diagram for this direction is shown in Fig.\,\ref{fig1}(b). The AFM phase (phase
III) is suppressed at $B_{\rm{N}} \sim 0.8$\,T, whereas the AFQ phase (phase II)
is suppressed at $B_{\rm{Q}} \sim 2$\,T. Both have been found to be continuous
by neutron scattering experiments \cite{Por16.1}. The continuous nature of the
transition at $B_{\rm{Q}}$ is also evidenced by the phase transition anomalies
in specific heat \cite{Ono13.1}, magnetostriction (fig.\,S1A,B),
and thermal expansion data (fig.\,S1C). The notion \cite{Cus12.1}
that the Fermi surface is large at $B > B_{\rm{N}}$ appears to have two
implications. Firstly, no further jump is to be expected at larger fields.
Indeed, it has been taken for granted that electron localization takes place
only once even in the case with multiple degrees of freedom. Secondly, the
quantum critical behavior at $B_{\rm{Q}}$ should be very different from that
near $B_{\rm{N}}$.

\begin{figure}[t]
{\centering
\subfigure[]{\includegraphics*[height=0.35\textwidth]{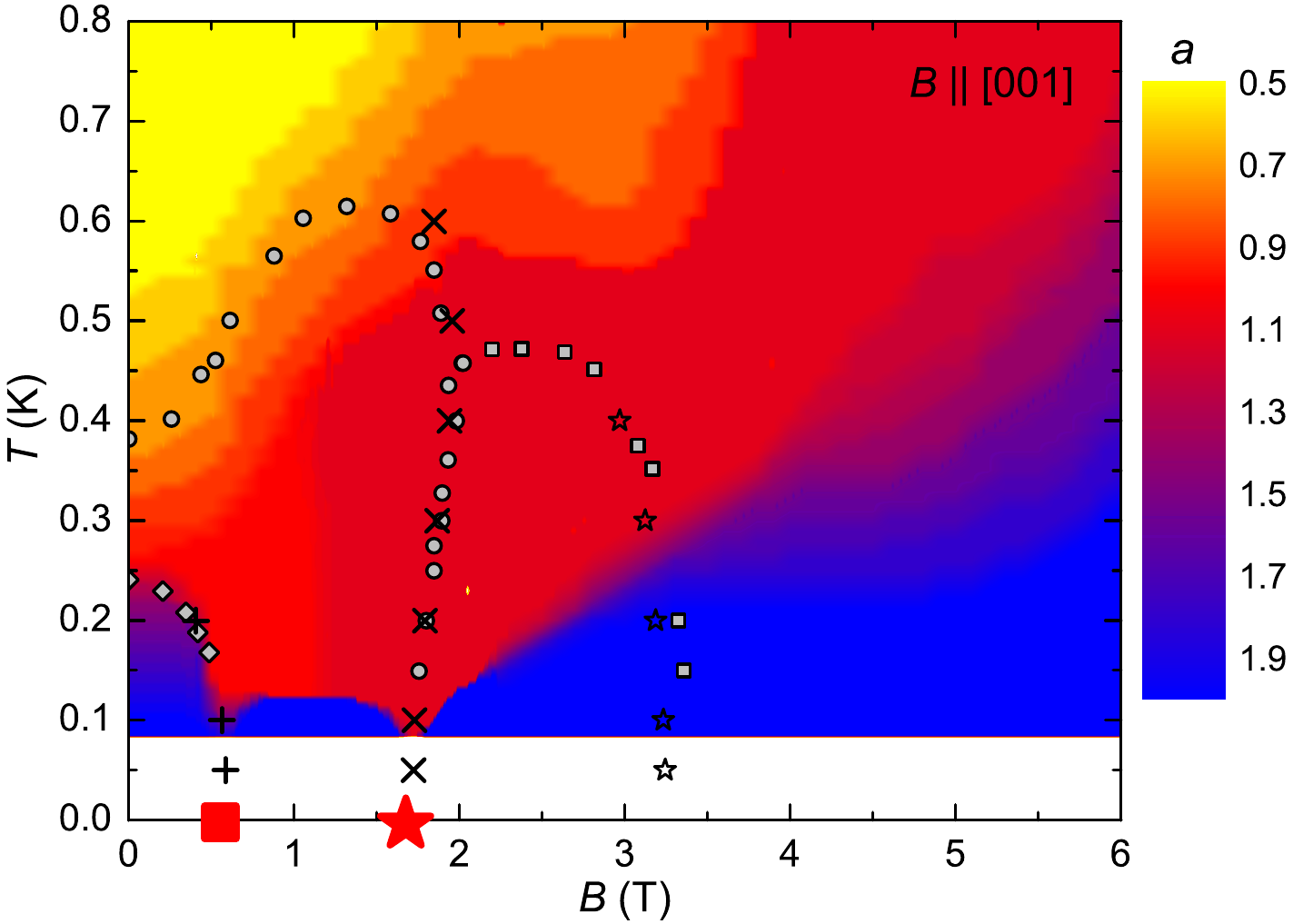}}\hspace{-1cm}
\subfigure[]{\includegraphics*[height=0.35\textwidth]{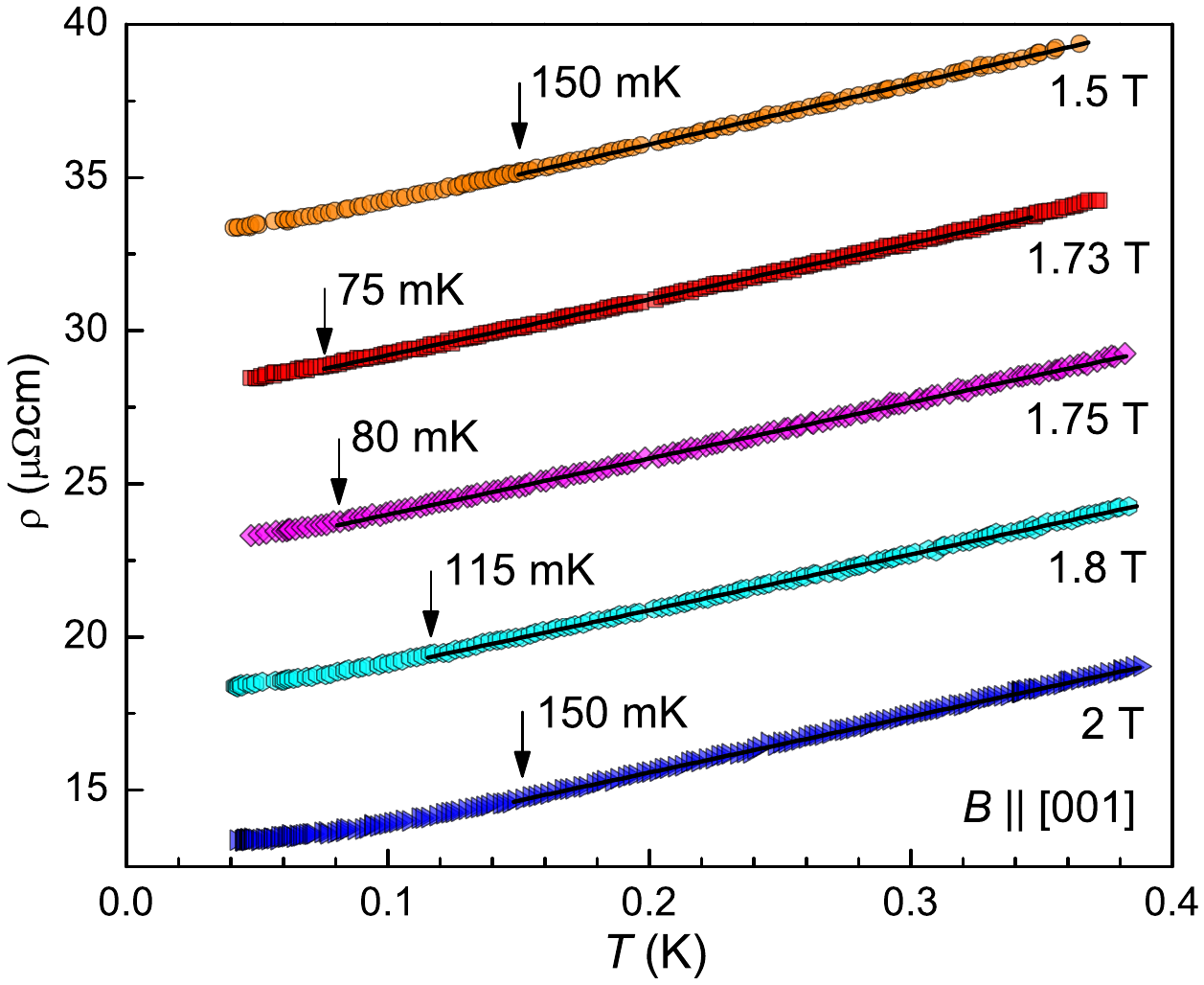}}
\subfigure[]{\includegraphics*[height=0.35\textwidth]{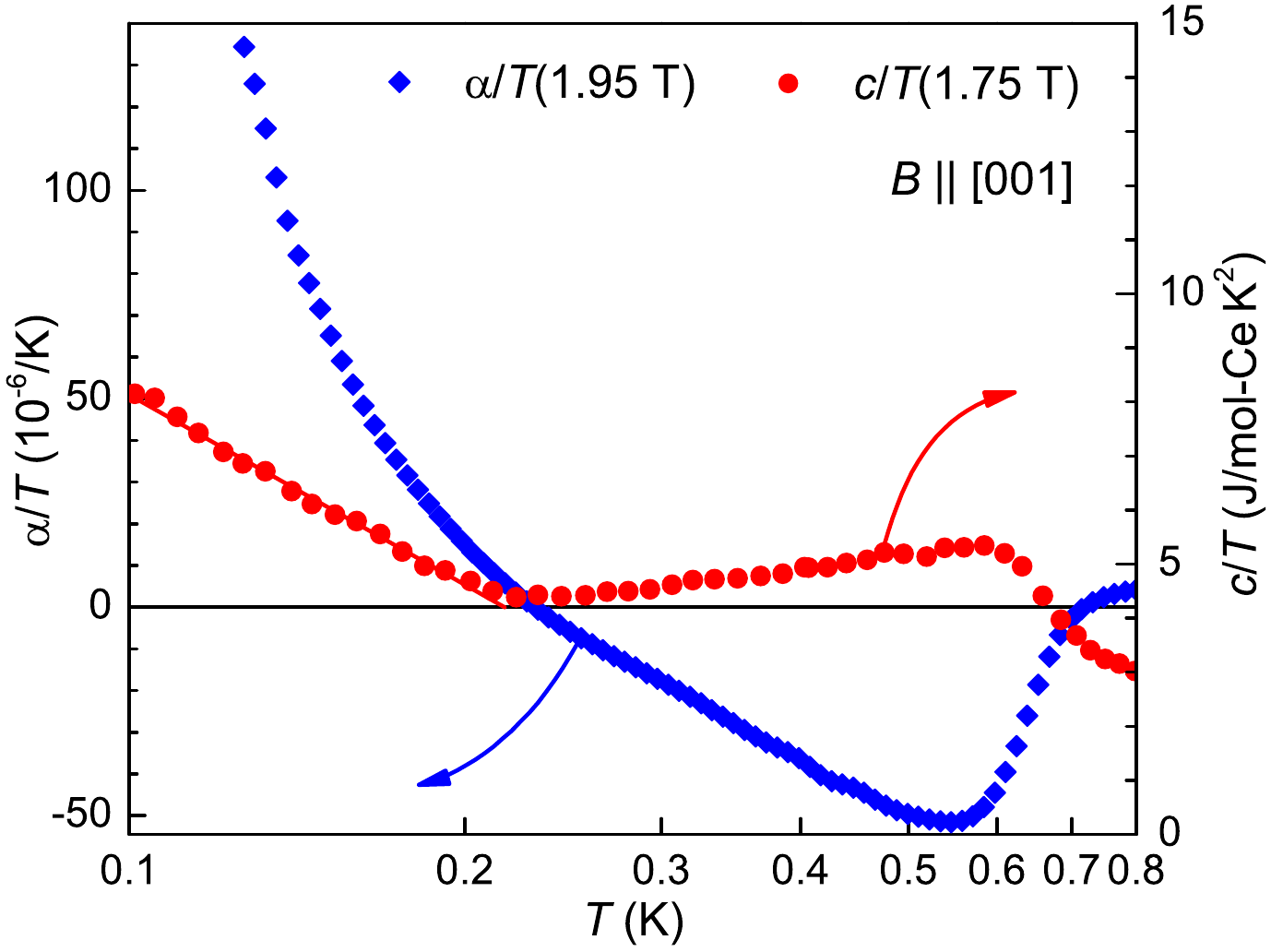}}\hspace{1.5cm} \subfigure[]{\includegraphics*[height=0.35\textwidth]{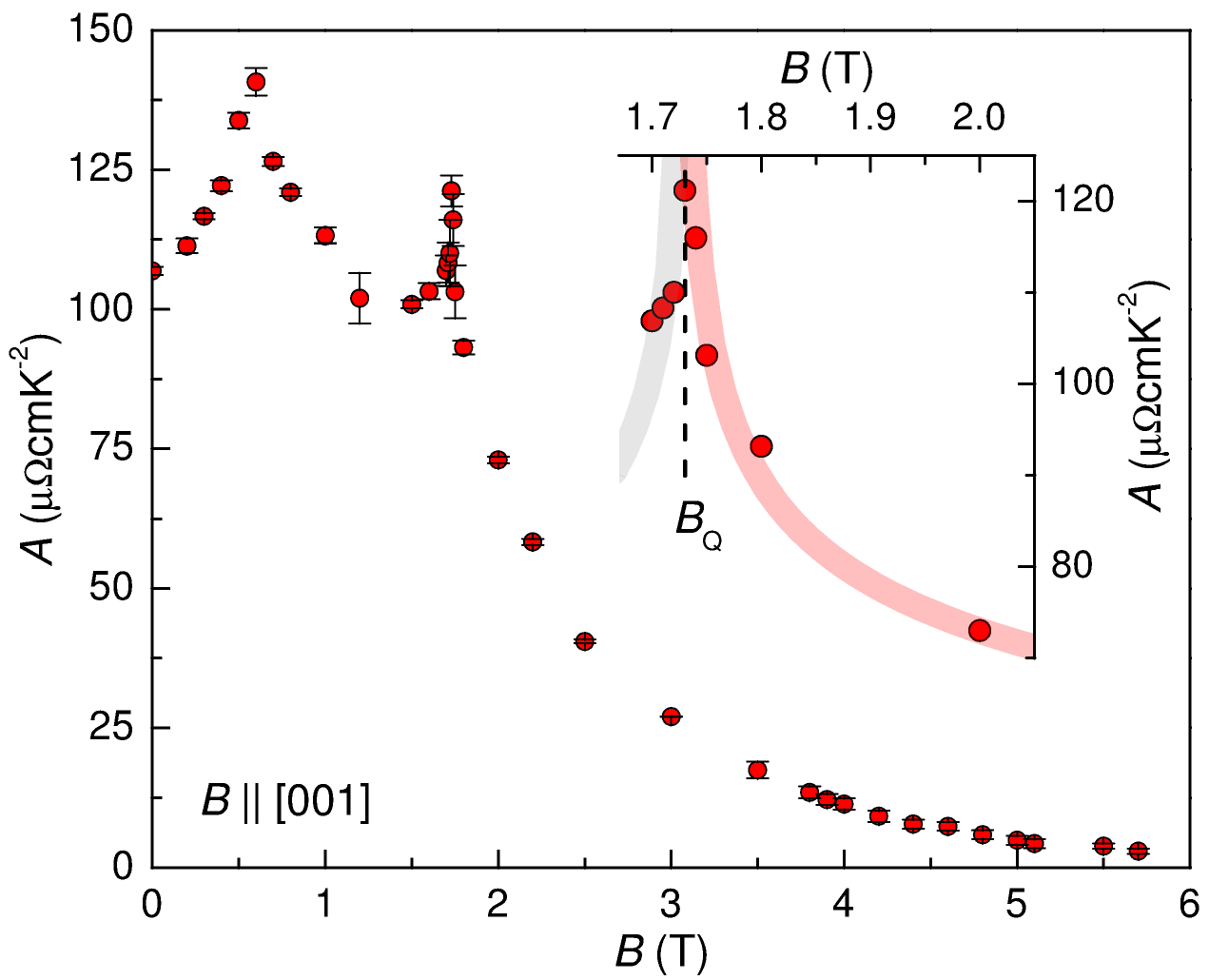}}}
\caption{\label{fig2}~{\bf Signatures of quantum criticality at the border of the AFQ phase in Ce$_3$Pd$_{20}$Si$_6$.} (a) Contour plot of the resistivity exponent $a$ of $\rho = \rho_0 +A'\cdot T^a$ in the temperature-magnetic field phase diagram for $B=\mu_0 H ||[0\,0\,1]$. To match the critical fields of our electrical resistivity sample, the fields of the phase transition lines (symbols) of Fig.\,\ref{fig1}(b) were slightly rescaled (Section S1). (b) Temperature-dependent electrical resistivity for selected magnetic fields $B=\mu_0 H || [0\,0\,1]$. Curves with fields above 1.5\,T are successively shifted downwards by 3\,$\mu\Omega$cm for better readability. The arrows indicate the temperatures down to which linear-in-$T$ behavior is observed, suggesting a critical field close to 1.73\,T. (c) Thermal expansion coefficient (left) and specific heat coefficient (right) vs.\ temperature near the respective critical fields $B_{\rm{Q}}$ [which are close to 1.95\,T for the thermal
expansion sample and 1.75\,T for the specific heat sample (Section S1)]. (d) $A$
coefficient of the FL part (see text) of the electrical resistivity vs.\ applied magnetic field $B=\mu_0 H$. The error bars represent standard deviations of the fit. The inset expands the field range around $B_{\rm{Q}}$, revealing the divergence of $A$. Lines are guides-to-the-eyes.
}
\end{figure}

Surprisingly, we find strange-metal behavior near $B_{\rm{Q}}$ that is 
strikingly similar to that near $B_{\rm{N}}$, as illustrated by the power-law
exponent $a$ of the temperature-dependent electrical resistivity ($\rho = \rho_0
+ A'\cdot T^a$) in the quantum critical fans anchored at $B_{\rm{Q}}$ and
$B_{\rm{N}}$, respectively (Fig.\,\ref{fig2}(a)). Indeed, at $B_{\rm{Q}}$, the
electrical resistivity $\rho$ is linear in temperature down to very low
temperatures (Fig.\,\ref{fig2}(b)), and the specific heat coefficient $c/T$
shows a logarithmic divergence (Fig.\,\ref{fig2}(c), right axis). In addition,
the thermal expansion coefficient $\alpha/T$ shows a stronger than logarithmic
divergence (Fig.\,\ref{fig2}(c), left axis), consistent with a diverging
Gr\"uneisen parameter $\Gamma \sim \alpha/c$. At fields away from $B_{\rm{Q}}$,
Fermi liquid (FL) behavior, with the form $\rho = \rho_0 + A\cdot T^2$, is
recovered in the electrical resistivity (Fig.\,\ref{fig2}(b), at temperatures
below the arrows). The $A$ coefficient, extracted from the respective FL regimes
(fig.\,S2), is strongly enhanced towards $B_{\rm{N}}$ and $B_{\rm{Q}}$
(Fig.\,\ref{fig2}(d)).

\begin{figure}[!ht]
\centering
\vspace{-2.5cm}

\subfigure[]{\includegraphics*[height=0.25\textwidth]{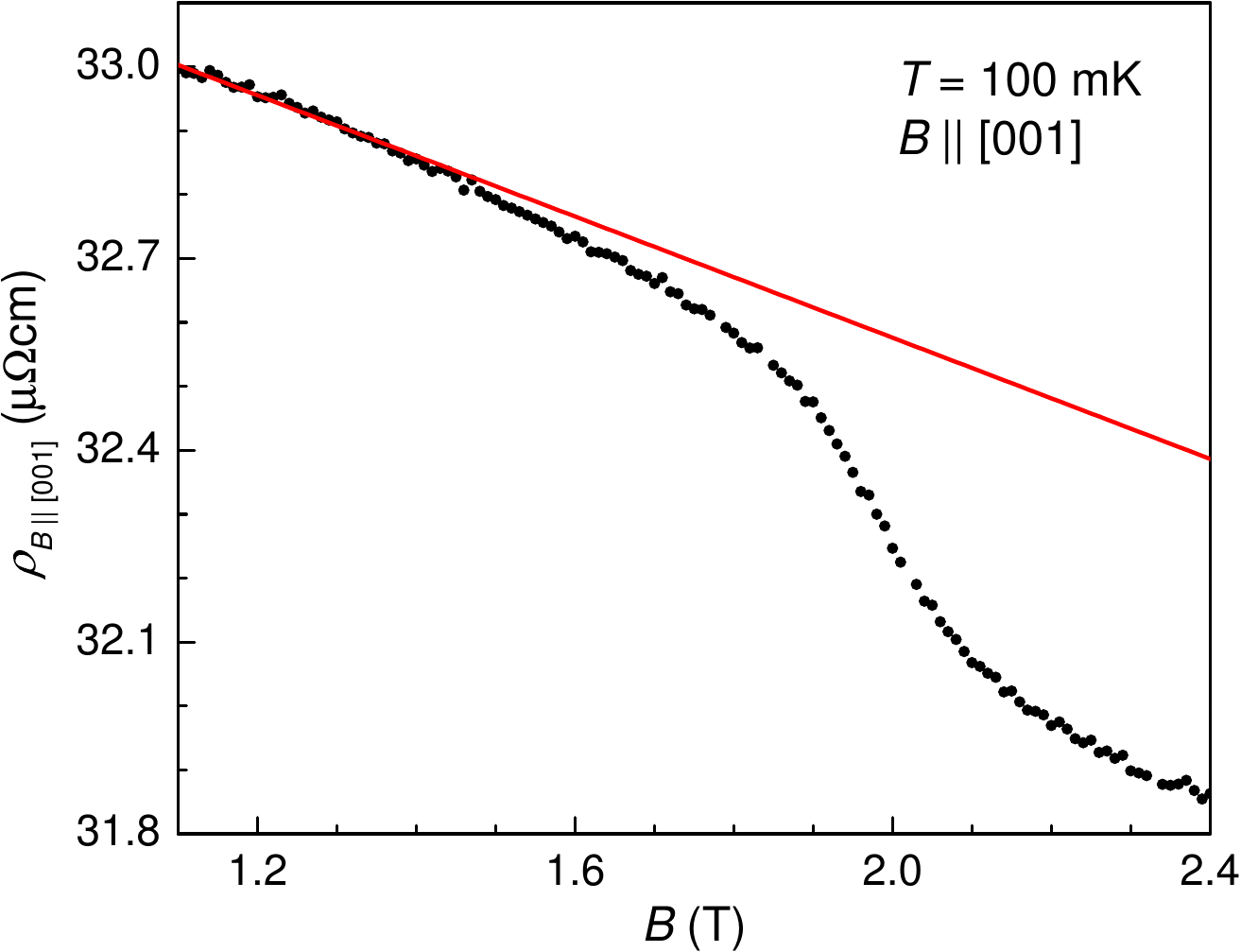}}\hspace{0.1cm}
\subfigure[]{\includegraphics*[height=0.25\textwidth]{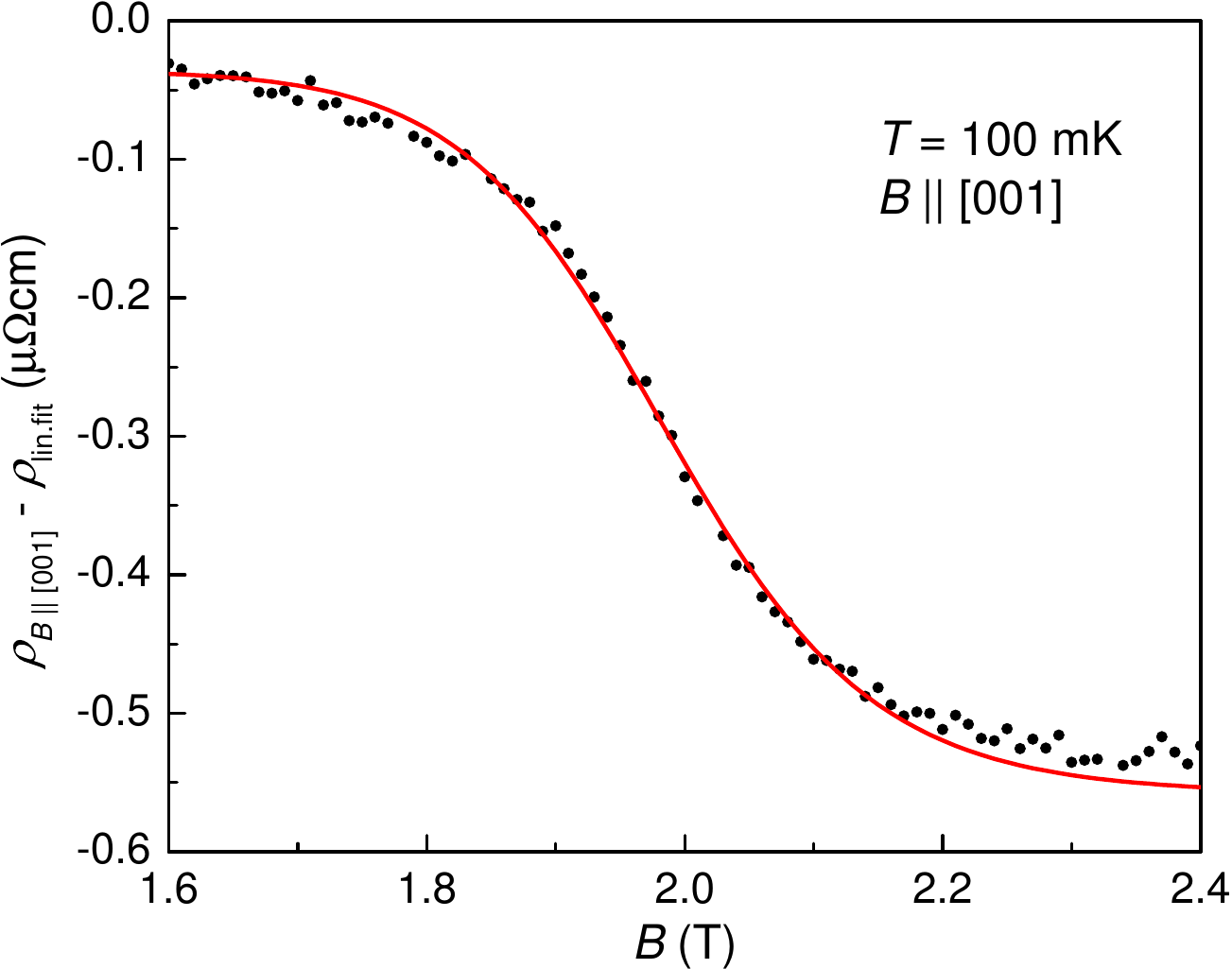}}\hspace{0.1cm}
\subfigure[]{\includegraphics*[height=0.25\textwidth]{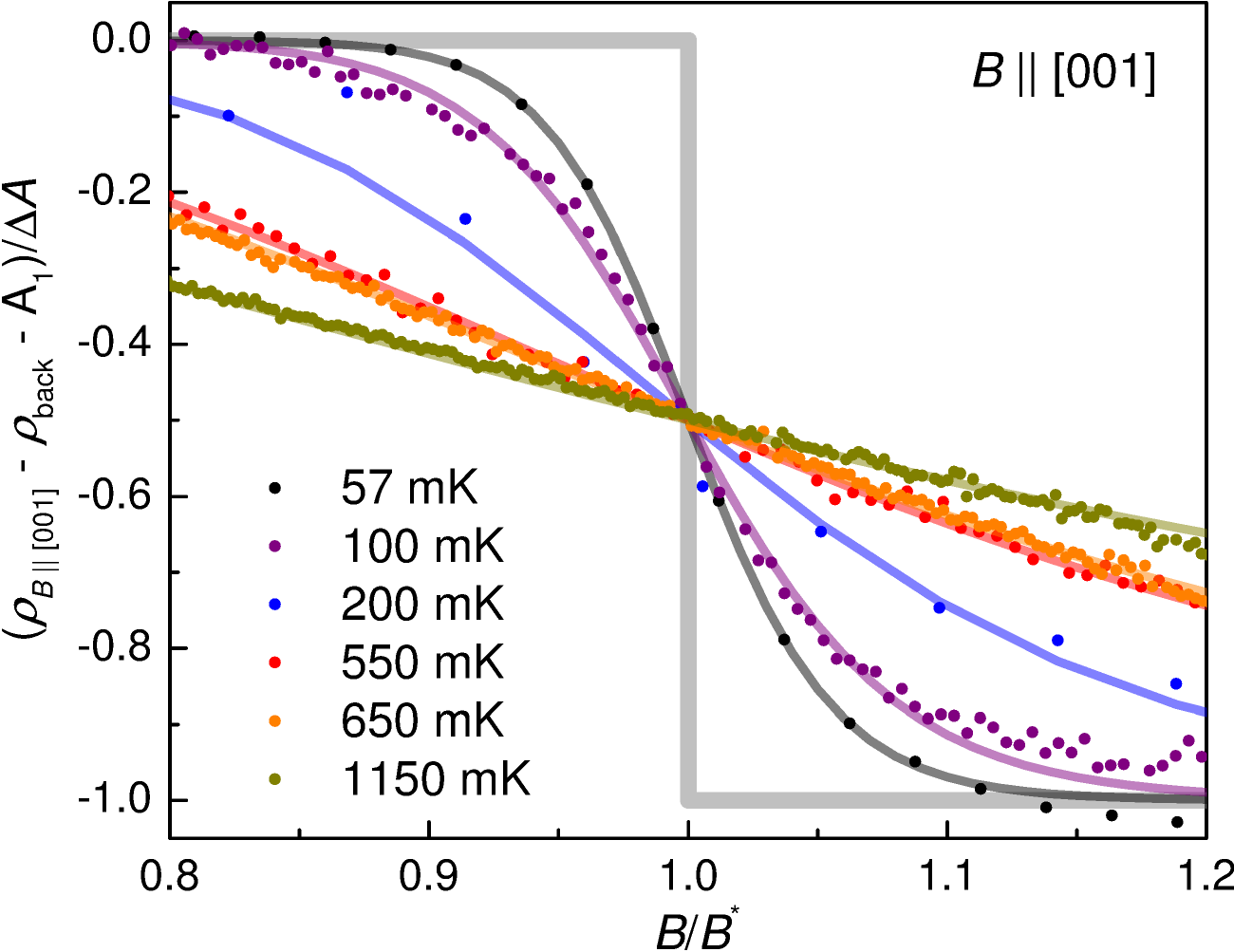}}
\subfigure[]{\includegraphics*[height=0.25\textwidth]{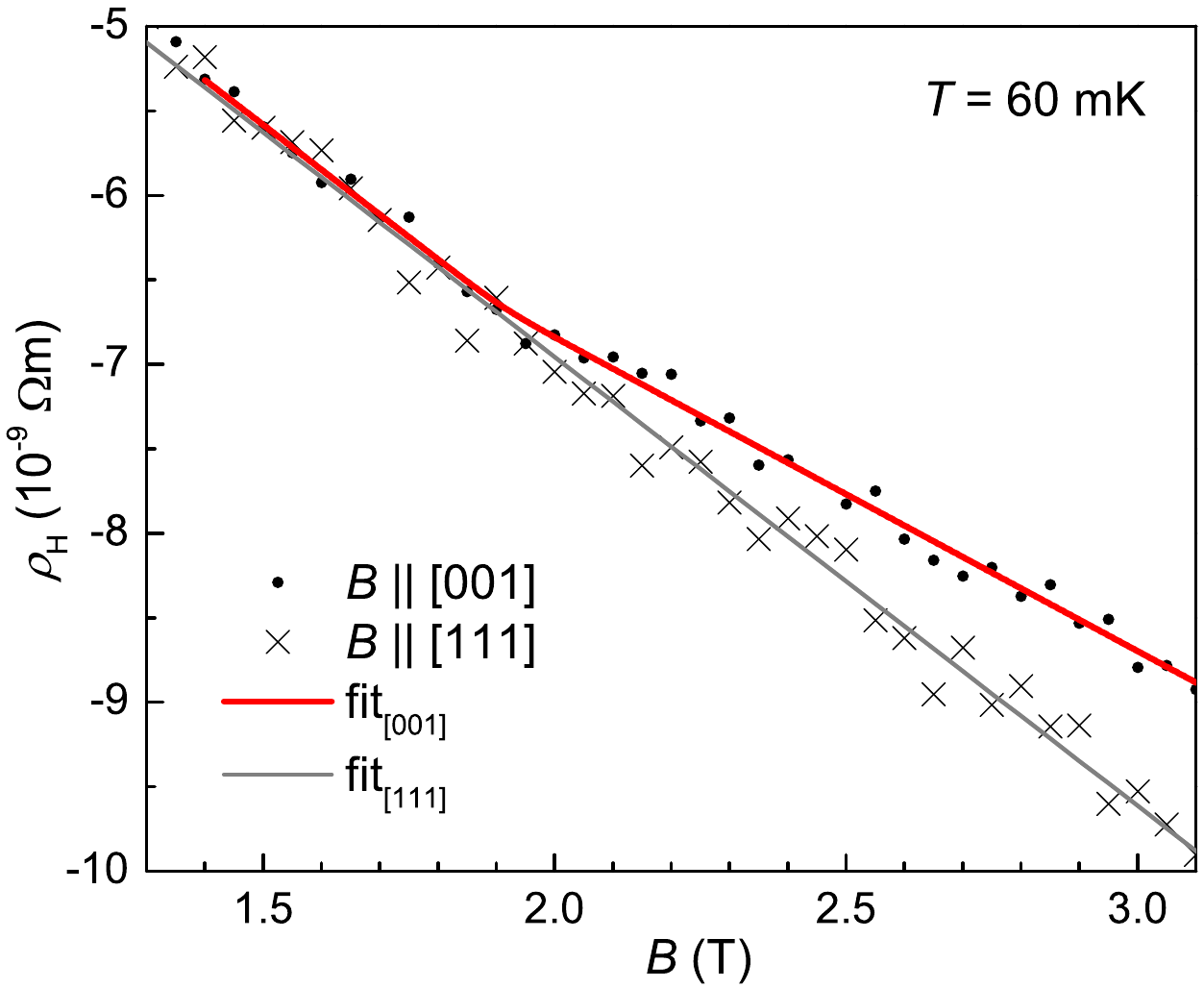}}\hspace{0.1cm}
\subfigure[]{\includegraphics*[height=0.25\textwidth]{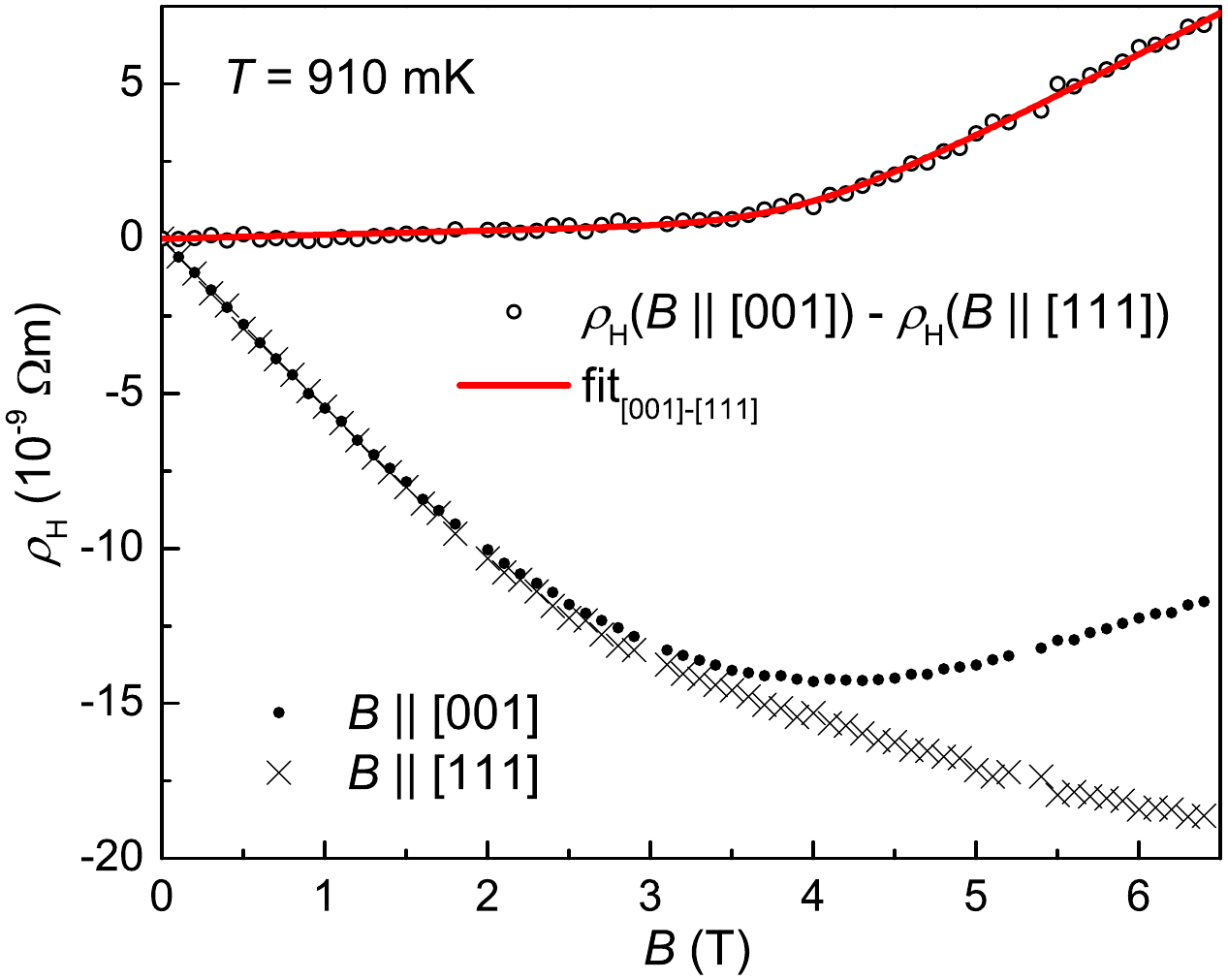}}\hspace{0.1cm}
\subfigure[]{\includegraphics*[height=0.25\textwidth]{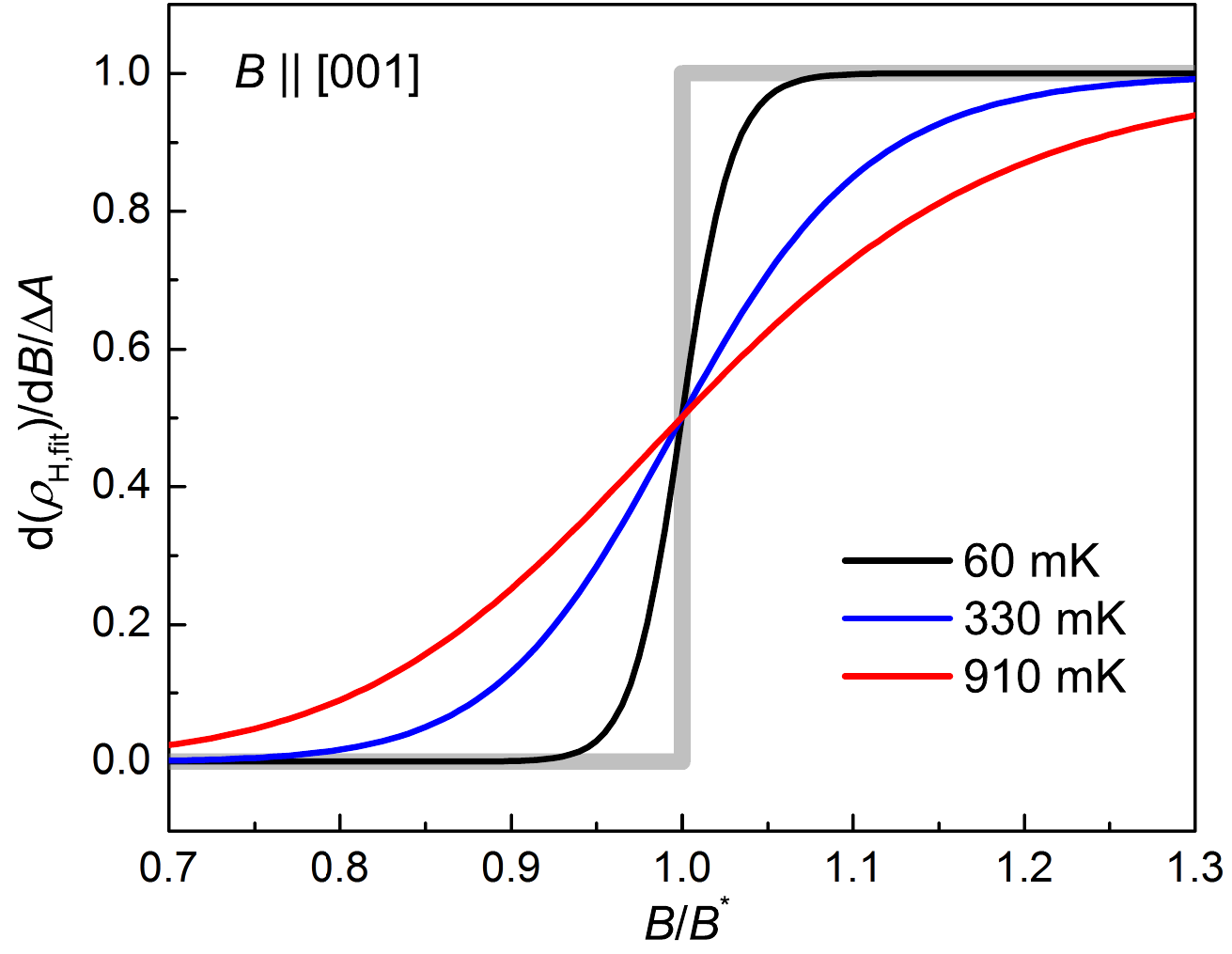}}
\subfigure[]{\includegraphics*[height=0.25\textwidth]{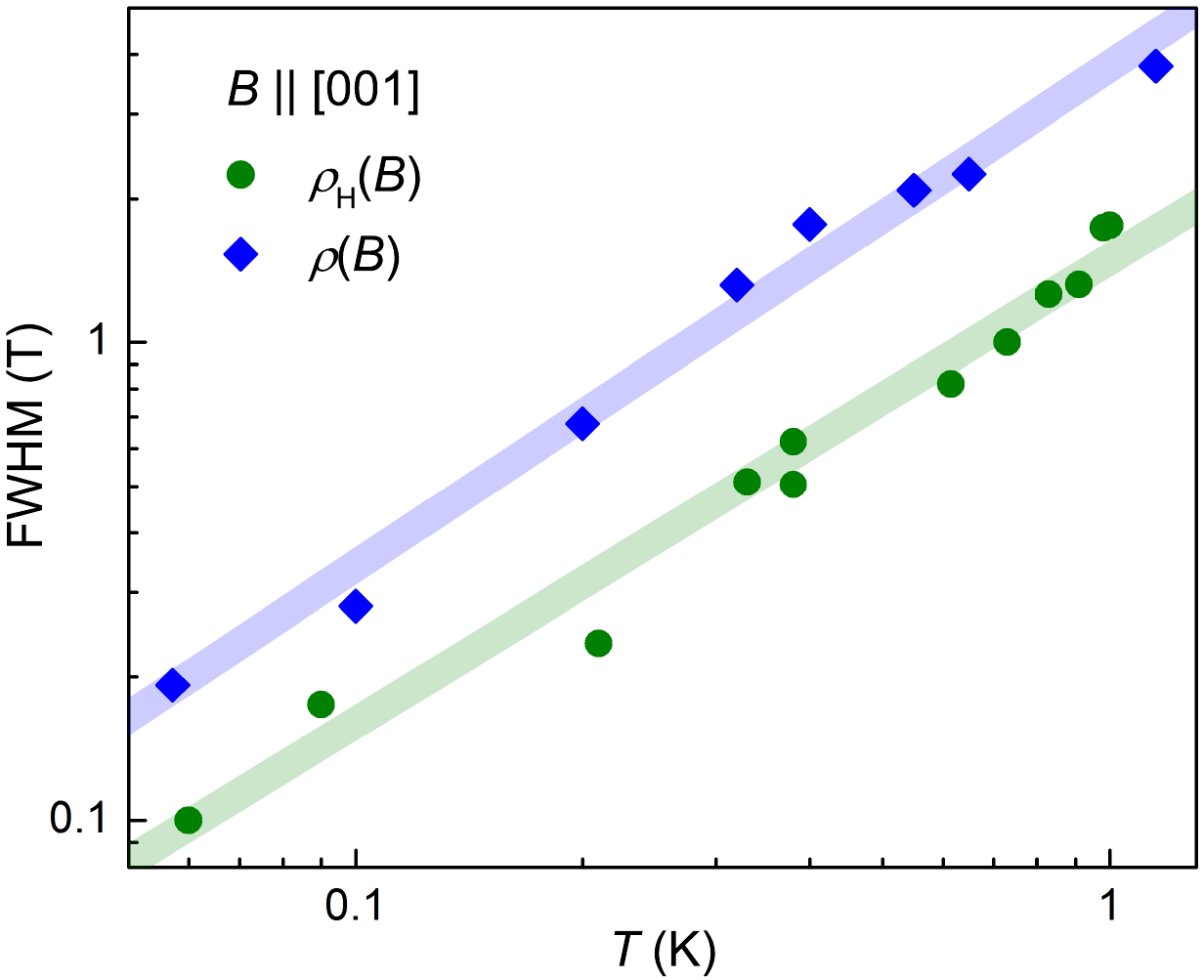}}\hspace{0.1cm}
\subfigure[]{\includegraphics*[height=0.25\textwidth]{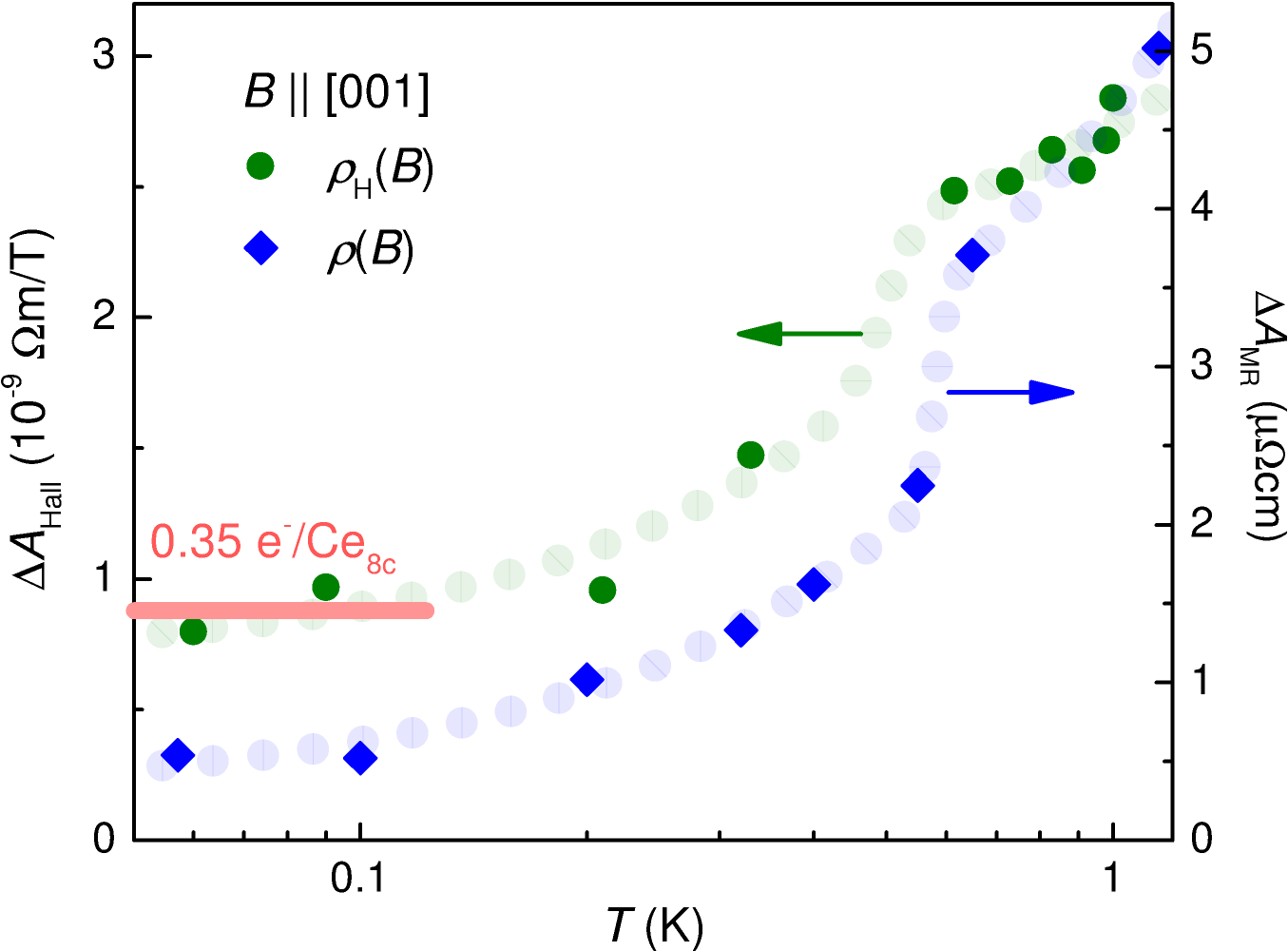}}
\caption{\label{fig3}~{{\bf Magnetotransport isotherms across the QCP at the
border of the AFQ phase in Ce$_3$Pd$_{20}$Si$_6$.} (a) Electrical
resistivity vs.\ magnetic field at 100\,mK. The red line represents a linear
background contribution. (b) Difference of electrical resistivity and the
background fit of (a). The red solid line represents a phenomenological
crossover fit (Section S3). (c) Selected scaled magnetoresistance
isotherms vs.\ scaled magnetic field (data points), together with the crossover
fits (full lines). An extended field range is shown in fig.\,S3. (d) Hall resistivity vs.\ magnetic field at 60\,mK, for two different field directions. The grey line represents a fit to the data for fields along $[1\,1\,1]$ for which no quantum criticality exists near 2\,T (Ref.\citenum{Mit10.1,Ono13.1}) and for which $\rho_{\rm{H}}$ is simply linear in $B$. The red line is a crossover fit (Section S3) to the data for
fields along $[0\,0\,1]$. Its low-field slope is fixed to the slope of the data
for fields along $[1\,1\,1]$. The full field range is shown in fig.\,S4.  (e) $\rho_{\rm{H}}(B)$ data at 910\,mK. Subtraction of the data for field along $[1\,1\,1]$ singles out the contribution due to the QCP at $B_{\rm{Q}}$ in the $\rho_{\rm{H}}(B||[0\,0\,1])$ data. (f) Selected scaled derivatives of the Hall resistivity crossover fits with respect to field vs.\ scaled magnetic field. (g) Full width at half maximum of the crossovers in magnetoresistance of (c) and the Hall resistivity derivatives of (f). The straight lines are best fits to ${\rm FHWM} \propto T^a$, with $a=1.05 \pm 0.05$ and $0.97 \pm 0.05$ for the magnetoresistance and Hall crossover, respectively. See also fig.\,S5.
 (h) Step heights of the magnetoresistance and Hall resistance crossovers. Indicated in red is the effective charge carrier concentration change, estimated using a spherical-Fermi-surface one-band approach. The thick grey lines in (c) and (f) correspond to extrapolations to $T=0$, where according to the FWHM the crossovers are sharp steps (``jumps").}}
\end{figure}

\begin{figure}[!ht]
\centering
\vspace{-2cm}

\subfigure[]{\includegraphics*[height=0.35\textwidth]{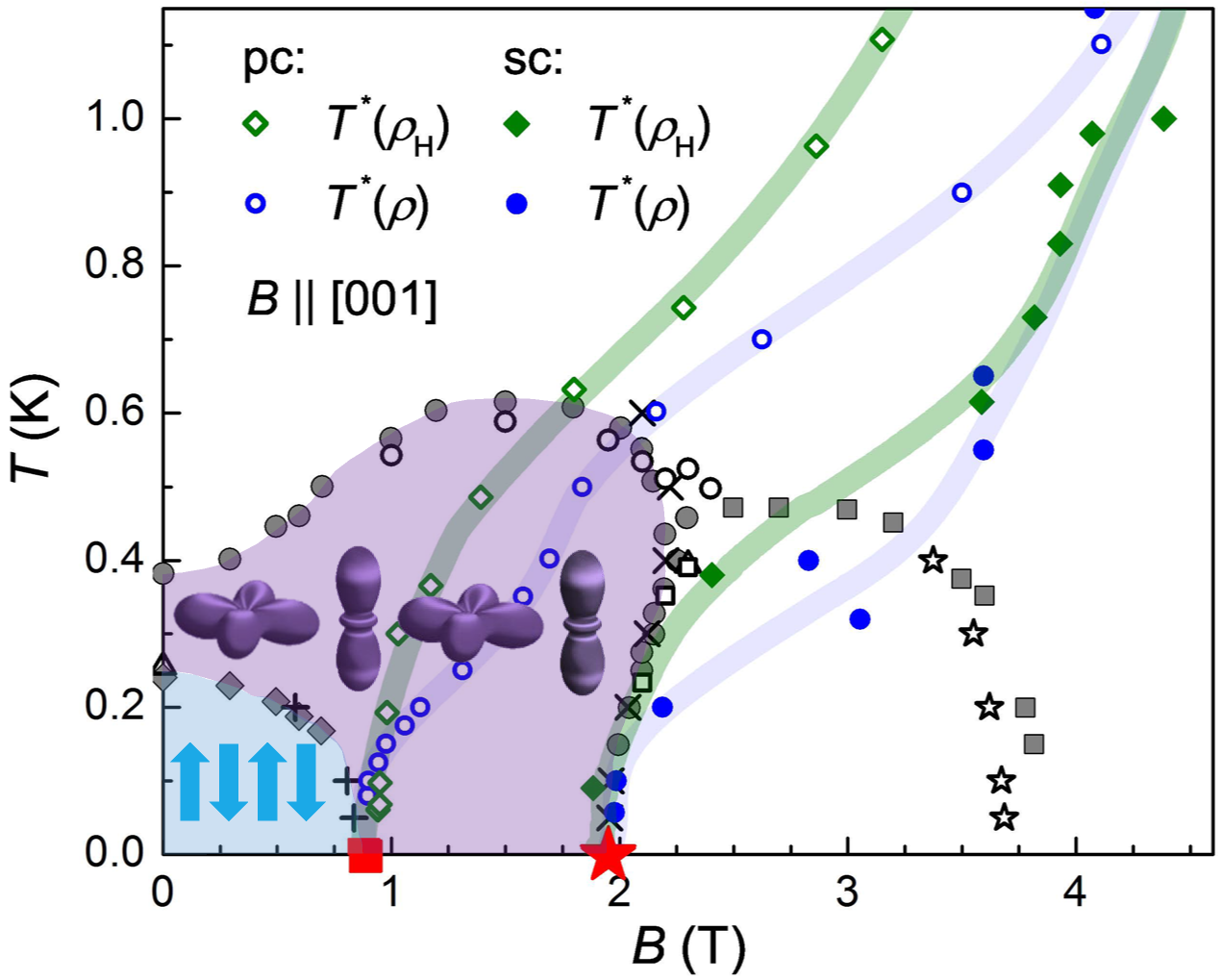}}\hspace{0.5cm}
\subfigure[]{\includegraphics*[height=0.35\textwidth]{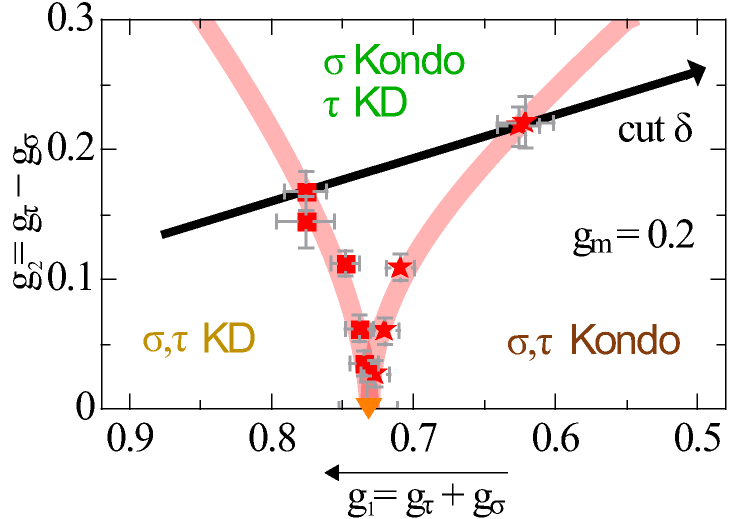}}\\[-0.1cm]
\subfigure[]{\includegraphics*[width=0.5\textwidth]{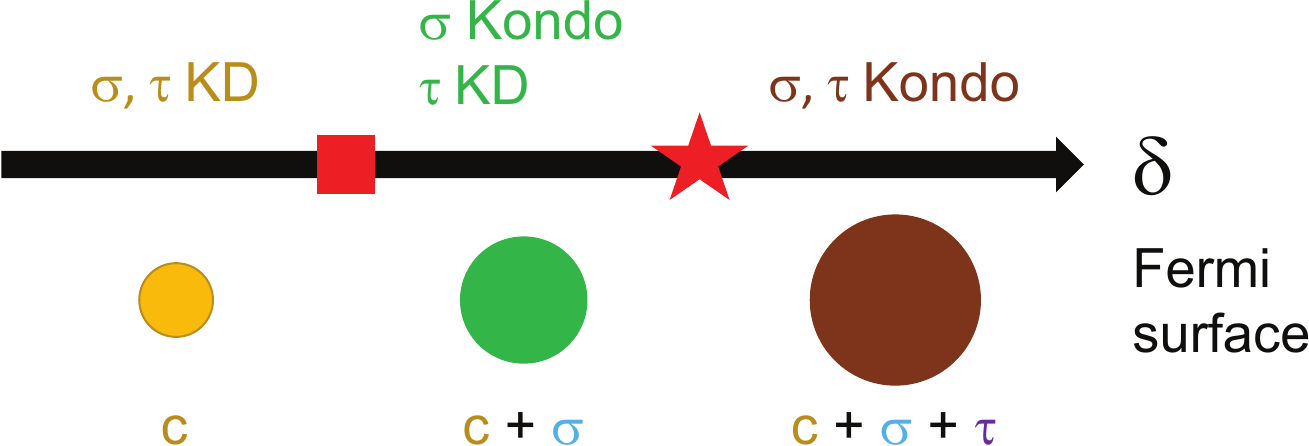}}\\
\subfigure[]{\includegraphics*[width=0.25\textwidth]{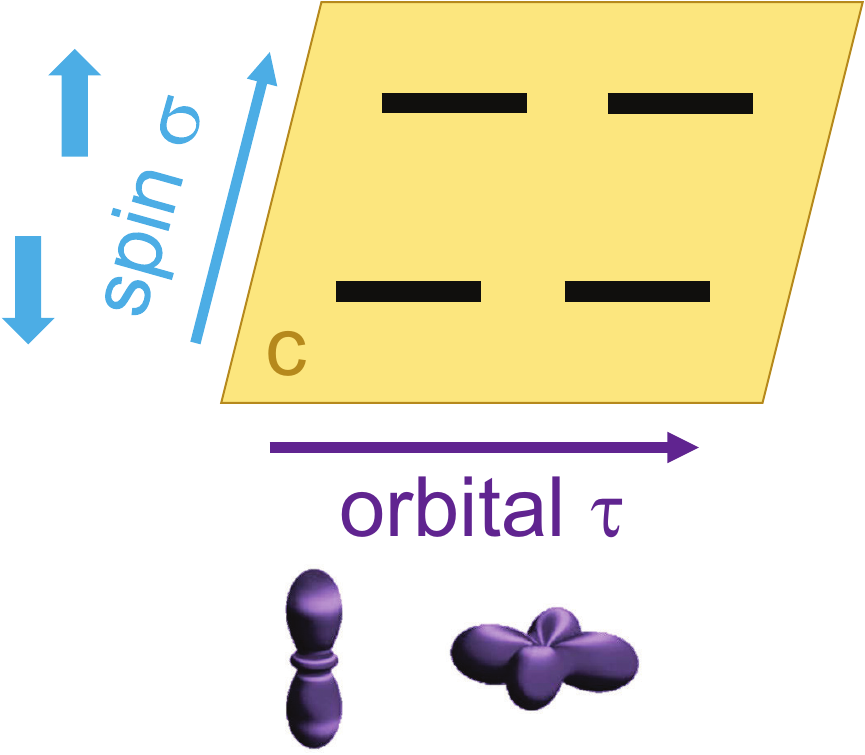}}\hspace{1.5cm}
\subfigure[]{\includegraphics*[width=0.25\textwidth]{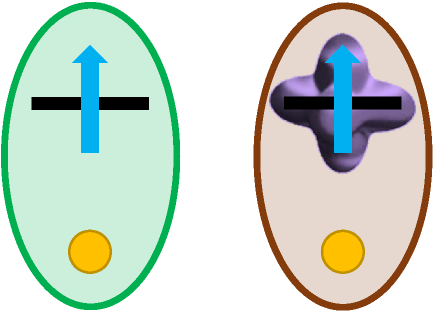}}
\newpage

\caption{\label{fig4}~{\bf Two-stage Kondo destruction in
Ce$_3$Pd$_{20}$Si$_6$.} (a) Experimental temperature--magnetic field phase
diagram from Fig.\,\ref{fig1}(b), with $T^{\ast}$ scales across which the Kondo
entanglement in the spin and orbital channel breaks up at two consecutive QCPs,
marked by the red square (at $B_{\rm{N}}$) and the red star (at $B_{\rm{Q}}$),
respectively. The $T^{\ast}$ scales at $B_{\rm{N}}$ are taken from Hall
resistivity [$T^{\ast}(\rho_{\rm{H}})$] and the magnetoresistance
[($T^{\ast}(\rho)$] measurements on a polycrystal (pc) \cite{Cus12.1}. The
corresponding $T^{\ast}$ scales at $B_{\rm{Q}}$, extracted from the
magnetotransport crossovers in Fig.\,\ref{fig3} for our transport single crystal
(sc), were slightly rescaled in $B$ to match the higher critical field of the
single crystals defining the phase boundaries (Section S1). The shaded
regions with the spin and orbital symbols visualize the AFM and AFQ phase,
respectively. (b) Theoretical phase diagram (at $T=0$) of the BFK model in
the $g_{1}$--$g_{2}$ plane. Red squares and stars mark the spin and orbital
Kondo destruction QCPs, respectively. The thick black arrow represents a generic
trajectory in the parameter space. The orange triangle represents the special
case $g_{2}=0$, where $g_{\tau}=g_{\sigma}$ and the two transitions occur
simultaneously. (c) Schematic of the sequential Kondo destruction
transitions, from a phase with Kondo destruction (KD) in both the spin
($\sigma$) and orbital ($\tau$) channel, via a phase where only the spin is
Kondo screened, to a phase with full Kondo screening. (d) Schematic of the
4-fold degeneracy of the $\Gamma_8$ ground state. (e) Sketches of the
Kondo entangled states with spin-only (left) and full Kondo entanglement
(right). The horizontal bars represent local degrees of freedom, the yellow
plane and circles the conduction electrons.
}
\end{figure}

To further characterize the behavior near $B_{\rm{Q}}$, we have measured the 
isothermal field-dependence of the electrical resistivity
(Fig.\,\ref{fig3}(a)-(c)) and the Hall resistivity (Fig.\,\ref{fig3}(d)-(f))
across this critical field. They reveal crossover signatures which can be
quantified following the procedures established previously
\cite{Pas04.1,Fri10.2,Cus12.1} (see also Section S3). The characteristic
parameters extracted from the analysis at each temperature are the full width at
half maximum FWHM of the crossover (Fig.\,\ref{fig3}(g)), the crossover height
$\Delta A$ (Fig.\,\ref{fig3}(h)), and the crossover field $B^{\ast}$ or,
equivalently, the field-dependent crossover temperature scale $T^{\ast}$
(Fig.\,\ref{fig4}(a)). The pure power-law behavior of the FWHM is seen as a
straight line in a double logarithmic plot (Fig.\,\ref{fig3}(g)); it
extrapolates to infinite sharpness in the zero-temperature limit, and thus a
jump in the Fermi surface. The power is 1 within error bars (see caption of
Fig.\,\ref{fig3}) for both the magnetoresistance and the Hall crossover at
$B_{\rm{Q}}$, similar to what was previously found for the quantum critical
point (QCP) at the border of the AFM phase in both Ce$_3$Pd$_{20}$Si$_6$
(Ref.\citenum{Cus12.1}) and YbRh$_2$Si$_2$ (Ref.\citenum{Pas04.1,Fri10.2}). Note
that, in the low-temperature limit, the change $\Delta n$ in the effective
charge carrier concentration across $B_{\rm{Q}}$, estimated using a simple
spherical-Fermi-surface one-band approach, is sizeable: it is about 0.35
electrons per Ce atom at the $8c$ site (Fig.\,\ref{fig3}(h)).

While a change in Fermi surface {\it per se} could come from a Lifshitz
transition, our observations near $B_{\rm{Q}}$ (and $B_{\rm{N}}$) are very
different. Lifshitz transitions for three-dimensional Fermi surfaces, as
observed in the high-field regime of YbRh$_2$Si$_2$ (Ref.\citenum{Pfa13.1}),
take place in the Fermi-liquid part of the phase diagram \cite{Geg06.1} and give
rise to only smooth evolutions of the Hall coefficient. Instead, strange-metal
behavior accompanied by a sizeable jump of the Fermi surface is the hallmark of
unconventional quantum criticality driven by Kondo destruction. 

The question, then, is how multiple stages of Kondo destruction may arise under
the tuning of a single control parameter. We consider a multipolar Kondo model
that contains a lattice of local moments with a 4-fold degeneracy (classified as
$\Gamma_8$ by the crystalline point group symmetry, see Section S4),
whose spin and orbital states are described by $\bm{\sigma}$ and $\bm{\tau}$,
respectively, and conduction electrons, $c_{{\bf k} \sigma \tau}$, as sketched
in Fig.\,\ref{fig4}(d). The $\Gamma_8$ moments are Kondo coupled to the conduction
electrons, and the coupling constants $J^{\kappa}_{\rm{K}}$ with $\kappa=\sigma,
\tau, m$, respectively, describe the interaction of $\bm{\sigma}$, $\bm{\tau}$,
and $\bm{\sigma \otimes \tau}$ with the conduction-electron counterparts. The
local moments also interact with each other by the RKKY exchange interactions
$I_{ij}^{\kappa}$ between sites $i$ and $j$ which, for the purpose of
computational feasibility, we have chosen to be of Ising type (Section
S5). In the extended dynamical mean field theory (Section S5),
this will be described in terms of the coupling between the local moments and
bosonic baths $\phi_{\kappa}$, with coupling constants $g_{\kappa}$. We are then
led to analyze the multipolar Bose-Fermi Kondo (BFK) model as an effective model for the Kondo lattice, which is described by the
Hamiltonian (see Section S5 for more details)
\begin{eqnarray}\label{H-BFK}
H_{\mathrm{BFK}} &=& H_{\mathrm{K}} + H_{\mathrm{BK}} +
H_{\mathrm{B0}}(\phi_{\sigma},\,\phi_{\tau},\,\phi_m )\quad , 
\end{eqnarray}
with $\quad H_{\mathrm{BK}} = g_{\sigma}\,\sigma^z\,\phi_{\sigma} +
g_{\tau}\,\tau^z\,\phi_{\tau} +g_m\,(\sigma^z \otimes \tau^z) \,\phi_m\quad.$ Here, $H_{\mathrm{K}}$ describes the Kondo coupling between the local
spin-orbital moments and the conduction electrons. In addition,
$H_{\mathrm{BK}}$ expresses the Bose-Kondo coupling between the local moments
and the bosonic baths whose dynamics are specified by $H_{\mathrm{B0}}$. 
For the pure (fermionic) Kondo part, our model corresponds to an exactly screened 
Kondo problem \cite{Hew97.1}, and is SU(4) symmetric when
 $J^{\kappa}_{\mathrm{K}}$ is the same for $\kappa=\sigma,\tau,m$.
Even when the SU(4) symmetry is broken, the system flows to
 the exactly screened (Fermi liquid) SU(4) Kondo fixed point \cite{Pan94,Hur04}.
 The model in the presence of bosonic Kondo couplings has not been studied before.
Based on what is known for the SU(2) Bose-Fermi Kondo model \cite{zhu02,zar02},
we expect that the overall phase diagram of the present model with 
different kinds of symmetries in the SU(4) space is captured by the calculations with
SU(4)-symmetric Kondo couplings and Ising-anisotropic bosonic couplings (see Supplementary Information,
Section S4).
We have
determined the zero-temperature phase diagram 
of this SU(4)-based Bose-Fermi Kondo model
via calculations using a
continuous-time quantum Monte Carlo method (Section S5).

The theoretical phase diagram is illustrated in Fig.\,\ref{fig4}(b), as a function
of $g_1 = g_{\tau} + g_{\sigma}$ and $g_2=g_{\tau}-g_{\sigma}$, for fixed
nonzero values of $g_m$ and $J_{\mathrm{K}}^{\kappa}$. Consider a generic
direction (cut $\delta$). In phase ``$\sigma$,\,$\tau$ Kondo'', both the spin
and orbital moments are Kondo entangled, which gives rise to an SU(4)-symmetric
electron fluid (Fig.\,\ref{fig4}(c),(e)\,right). Upon moving towards the left
(against direction of arrow $\delta$), this state first undergoes the
destruction of Kondo effect in the orbital sector at one QCP (stars in
Fig.\,\ref{fig4}(b),(c)). This drives the system into a phase in which only the spin
moments form a Kondo singlet with the conduction electrons (phase ``$\sigma$
Kondo, $\tau$ KD'' in Fig.\,\ref{fig4}(b),(c), Fig.\,\ref{fig4}(e)\,left). It then, at
the next QCP (squares in Fig.\,\ref{fig4}(b),(c)), experiences the destruction of
Kondo effect in the spin sector, leading to a fully Kondo destroyed state (phase
``$\sigma$,\,$\tau$ KD'' in Fig.\,\ref{fig4}(b),(c)). Consequently, in a multipolar
Kondo lattice system, there will be two distinct QCPs associated with a sequence
of Kondo destructions. At each of the QCPs, the Fermi surface undergoes a sudden
reconstruction (circles in Fig.\,\ref{fig4}(c)), which explains the jumps inferred
from the Hall coefficient and magnetoresistance data. For a single-band
jellium-like electronic fluid, our theory implies an integer jump of the
electron count at each QCP. Any real material would, however, show deviations
from this equality, as also seen here (see above). The sharp statement is that a
jump in the electron count and Fermi surfaces must be manifested in the
extrapolated zero-temperature limit of the Hall crossover, as we have
demonstrated. We stress that an applied magnetic field is
expected to weaken magnetic order more rapidly ($\sim B$) than the Kondo
processes ($\sim B^2$); related considerations apply to the quadrupolar sector.
Thus, the sequential Kondo destruction happens upon decreasing the magnetic
field, i.e., from right to left in the experimental phase diagram
(Fig.\,\ref{fig4}(a)).

We have thus demonstrated that, in spite of the genuine intermixing of the two
degrees of freedom in the many-body dynamics, a remarkable separation of their
fingerprints occurs in the singular physics of quantum criticality: The
magnetic-field tuning realizes two stages of quantum phase transitions, which
are respectively dictated by the Kondo destruction of the spin and orbital
sectors.\\


\noindent{DISCUSSION}\\
To put this finding in perspective, we recall that in spin-only systems,
experiments have provided extensive evidence for Kondo destruction in AFM heavy
fermion compounds \cite{Sch00.1,Pas04.1,Shi05.1,Fri10.2,Cus12.1,Mun13.1}. From
studying a spin-orbital heavy fermion system, we have shown that Kondo
destruction is a general phenomenon and may also occur if degrees of freedom
other than spin decouple from the conduction electrons. This demonstrates Kondo
destruction as a general framework for both beyond-Landau quantum criticality
and the electron localization-delocalization transition in metallic heavy
fermion systems. Our analysis of the multipolar degrees of freedom also relates
to the purely orbital case, as realized for instance in the Pr-based heavy
fermion systems PrV$_2$Al$_{20}$ (Ref.\citenum{Shi15.1}) and PrIr$_2$Zn$_{20}$
(Ref.\citenum{Oni16.1}). These materials show unusual multipolar quantum
criticality, though Kondo destruction has not yet been explored. Future studies
may reveal whether electron localization occurs in these orbital-only heavy
fermion systems as well, and contributes to nucleating phases
\cite{Myd11.1,McC13.1} -- including unconventional superconductivity
\cite{Bau02.1,Mat12.1}.

More generally, we have demonstrated that strange-metal properties occur at each
stage of the electron localization transition. This finding connects well with
other classes of strongly correlated systems in which strange-metal behavior has
also been linked to electron localization. In the high-$T_{\rm{c}}$ cuprate
superconductors, electron localization as suggested by a pronounced change of
the Fermi surface \cite{Bad16.1,Bal03.1} and a divergence of the charge carrier
mass \cite{Ram15.1} appears near the hole doping for optimal superconductivity,
where strange-metal properties arise. In organic systems, electron localization
has also been evidenced in connection with strange-metal behavior and optimal
superconductivity \cite{Oik15.1}. In the graphene superlattices with a
magic-angle twist, whose electronic states may also satisfy an SU(4) symmetry
from the combination of the spin and valley degrees of freedom, transport and
quantum oscillation measurements \cite{Cao18.1} have implicated a ``small" Fermi
surface of the charge carriers doped into a Mott insulator, thereby raising the
possibility of an electron localization-delocalization transition underlying the
superconductivity. As such, our work provides new understandings on the
breakdown of the textbook description of electrons in solids and points to
electron localization as a robust organizing principle for strange-metal
behavior and, by extension, high-temperature superconductivity.

Our system contains strongly correlated and entwined degrees of freedom; the
crystalline symmetry dictates the strong intermixing of the spin and orbital 
quantum numbers. Yet, near each of the two QCPs, there is a clear selection of
the orbital or spin channel that drives the quantum critical singularity. This
remarkable simplicity, developed out of the intricate interplay among the
multiple degrees of freedom, represents a new insight into the physics of
complex electron fluids. This new understanding may also impact on strongly
correlated systems beyond the realm of materials such as mesoscopic structures
\cite{Kel14.1} and quantum atomic fluids \cite{Nak16.2,Neu06.1} where
localization-delocalization transitions may also play an important role.
Finally, the sequential localization we have advanced may be viewed as
selectively coupling only part of the system to an environment. This notion
relates to ideas for reduced dephasing within a logical subspace \cite{Fri17.1},
and may as such inspire new settings for quantum technology.

\noindent Materials and methods are described in the Supplementary Information.


\noindent{\bf Acknowledgements:}
The authors wish to thank D.\ Joshi for his contribution to the crystal growth,
R.\ Dumas from Quantum Design for contributing to the heat capacity
measurements, T.\ Sakakibara for sharing data of Ref.\,\citenum{Mit10.1} with
us, L.\ B\"uhler for graphical design, and E.\ Abrahams, S.\ Kirchner, D.\
Natelson, A.\ Nevidomskyy, T.\ Park, and S.\ Wirth for fruitful discussions. The work in Vienna was funded by the Austrian Science Fund (P29296-N27 and DK
W1243), the European Research Council (Advanced Grant 227378), and the US Army Research Office (ARO-W911NF-14-1-0496). The work at Rice was in part supported by the National Science Fundation (DMR-1920740) and the Robert A. Welch Foundation (C-1411) (A.C., E.M.N., C.-C.L., Q.S.), the Army Research Office (W911NF-14-1-0525) and a Smalley Postdoctoral Fellowship at the Rice Center for Quantum Materials (H.-H.L.), and the Big-Data Private-Cloud Research Cyberinfrastructure MRI Award funded by NSF (CNS-1338099) and by an IBM Shared University Research (SUR) Award. V.M.\ was supported by the FAPERJ (201.755/2015), R.Y.\ by the National Science Foundation of China (11374361 and 11674392) and the Ministry of Science and Technology of China (National Program on Key Research, 2016YFA0300504), 
K.I.\ by the National Science Foundation (Grant No.\, DMR-1508122),
and A.M.S.\ by the SA-NRF(93549) and the UJ-FRC/URC. 
Q.S.\ acknowledges the hospitality of the Aspen Center for Physics (NSF, PHY-1607611) 
and University of California at Berkeley.



\end{document}